\documentstyle[manuscript,aps]{revtex}

\newcommand{\beq}{\begin{equation}}
\newcommand{\eeq}{\end{equation}}
\newcommand{\beqa}{\begin{eqnarray}}
\newcommand{\eeqa}{\end{eqnarray}}
\newcommand{\beqar}{\begin{eqnarray*}}
\newcommand{\eeqar}{\end{eqnarray*}}

\def \T {{\bf T_A}}
\def \la {\langle}
\def \ra {\rangle}

\def \x {{\bf x}}
\def \y {{\bf y}}
\def \p {{\bf p}}
\def \px {{\bf P_x}}
\def \py {{\bf P_y}}
\def \up {\uparrow}
\def \down {\downarrow}
\def \G {{\bf G}}
\begin{document}
\input epsf
\title{\Large  
Measurement of Time-of-Arrival in Quantum Mechanics
}

\author{ {Y. Aharonov$^{(a)}$, 
J. Oppenheim$^{(b)}$,\footnote{\it jono@physics.ubc.ca}
S. Popescu$^{(c)}$,\footnote{\it sp230@newton.cam.ac.uk}   
B. Reznik$^{(d)}$,\footnote{\it reznik@t6-serv.lanl.gov}
and W. G. Unruh$^{(b)}$}\footnote{\it unruh@physics.ubc.ca \\
LAUR-96-4753}  
{\ } \\
{\footnotesize
(a) { \it   School of Physics and Astronomy, Tel Aviv
University, Tel Aviv 69978, Israel, and  Department of Physics, 
University of South Carolina, Columbia, SC 29208.}\\
(b) { \it   Department of Physics,  University of British
Columbia,
 6224 Agricultural Rd. Vancouver, B.C., Canada
V6T1Z1}\\
(c) { \it Isaac Newton Institute, Cambridge University, 
 20 Clarkson Rd., Cambridge, CB3 OEH, U.K.
and BRIMS, Hewlett-Packard Labs., Bristol BS12 6QZ, U.K}\\ 
(d) {\it  Theoretical Division, T-6, MS B288, 
Los Alamos National Laboratory, Los Alamos, NM, 87545}
}}
\maketitle 
\begin{abstract}
{ 
It is argued that the time-of-arrival cannot be precisely
defined and  measured in quantum mechanics. 
By constructing explicit toy models of a measurement, we 
show  that for a free particle 
it cannot be measured more accurately then 
$\Delta t_A \sim 1/E_k$, where $E_k$ is the initial 
kinetic energy of the particle. With  a better accuracy,  
particles reflect off the measuring device, and the resulting 
probability distribution becomes  distorted.
It is shown that a time-of-arrival operator cannot exist, and that 
approximate time-of-arrival operators do not
correspond to the measurements considered here.
} 

\end{abstract}

\newpage

\section{Introduction}

Consider a beam of free particles, upon which a 
measurement is performed to determine the time-of-arrival 
to $x=x_A$. The time-of-arrival can be 
recorded by a clock situated at $x=x_A$ which switches off when the particle
reaches it.
%All that is needed in this measurement 
%is to switch the running clock off when
%the particle reaches the clock. 
In classical mechanics we could, in principle, achieve this 
with the smallest non-vanishing interaction between the particle
and the clock, and 
hence measure the time-of-arrival with arbitrary accuracy. 

In classical mechanics there is also another
indirect method to measure the time-of-arrival. 
First invert the equation of motion of the
particle and obtain the time in terms of the location 
and momentum, $T_A(x(t), p(t), x_A)$. 
This function can be determined
at {\it any time} $t$, either by a simultaneous
measurement of $x(t)$ and $p(t)$ and evaluation of $T_A$, 
or by a direct coupling to $T_A(x(t), p(t), x_A)$.
  
These two different methods, namely, the direct measurement,
and indirect measurement, are classically equivalent. 
They give rise to the same classical time-of-arrival.
They are not equivalent however,  in quantum mechanics 
%\cite{srinivas}.

In quantum mechanics 
the corresponding operator ${\bf T_A}( \x(t), \p(t), x_A)$,
if well defined, can in principle 
be measured to any accuracy. On the other hand, it has been argued by Allcock\cite{allcock2}, that a 
direct measurement cannot determine    
the time-of-arrival of free particles to any accuracy.
In section IIB, we argue that Allcock's arguments are not sufficient
to limit the accuracy of time-of-arrival measurements. One needs to
consider models with physical clocks. Using these models, we shall argue 
that the accuracy of time-of-arrival measurements cannot be better than 
\beq
\Delta t_A > 1/E_k,
\label{limitfirst}
\eeq 
where $E_k$ is the initial kinetic energy of the particle and we use units
with $\hbar=1$.
The basic reason is that, unlike a classical mechanical
clock, in quantum mechanics the uncertainty in the clock's
energy grows when its accuracy improves \cite{salecker}.
We find that particles with initial kinetic energy
$E_k$ are reflected without switching off a clock if this clock 
is set to record the
time-of-arrival with accuracy better than in 
eq. (\ref{limitfirst}). (The occurrence of a similar
phenomenon is well known in optics as 
an impedance miss-match which causes reflection in wave guides.)
Furthermore, for the small 
fraction of the ensemble that does 
manage to turn off the clock, the resulting 
probability distribution becomes distorted.
A detailed discussion of direct time-of-arrival measurements 
is given in Section II.

Still, one can imagine an indirect determination
of arrival time as described above,
by a measurement of some regularized time-of-arrival operator 
$\T (\x(t), \p(t), x_A)$ \cite{rovelli}.
An obvious requirement of $\T$ is that 
it is a constant of motion; i.e., the time-of-arrival 
cannot change in time.
As we shall show in Section III, a Hermitian time-of-arrival
operator, with a continuous spectrum, 
can satisfy this requirement 
only for systems with an unbounded Hamiltonian.
This difficulty can however by circumvented by ``projecting out''
the singularity at $p=0$ and by using only measurements of $\T$
which do not cause a ``shift'' of the energy towards
the ground state.
Nevertheless, unlike the classical case, 
in quantum mechanics the result of 
such a measurement may have nothing to do with 
the time-of-arrival to $x=x_A$. 
As is argued in Section IV, since $\T$ can be measured 
with arbitrary accuracy
it does not correspond to the result obtained by the  
direct measurement discussed in Section II. 
We conclude in Section V. 
by a discussion of the main results.
An explicit calculation of the clock's final 
probability distribution is given in the  Appendix.

\section {Measurement of time-of-arrival}

In this section we consider 
toy models of a measurement 
of time-of-arrival.
To begin with, assume that a beam of particles interacts 
with a detector that is located at $x=0$ and 
is coupled to a clock. Initially, as the beam is prepared, the 
clock is set to show $t=0$. Our purpose is to design 
a particular set-up such that as a particle
crosses the point $x=0$ the detector stops the clock.
Since the masses of the particle detector and the
 clock are unlimited we can ignore the uncertainty
in the position of the measurement device.
We shall consider four models. The first model
describes a direct interaction of the particle
with the clock. 
In the second model, the particle is detected by 
a two-level detector, which turns the clock off.
To avoid the reflection due to ``impedance miss-match'', 
we look next at the possibility of boosting the 
energy of the particle
in order to turn off the clock.  
%In order to avoid reflection due to ``impedance miss match'',
We shall also consider the case of a 
"smeared" interaction, and conclude with a general 
discussion.
 
\subsection{Measurement with a clock}

The simplest model which describes a direct interaction
of a particle and a clock \cite{peres}, without additional
"detector" degrees of freedom,
is described by the Hamiltonian
\beq
H = {1\over 2m} {\px}^2 + \theta(-\x) {\py}.
\label{model1clock}
\eeq
Here, the particle's motion is confined to one 
spatial dimension, $x$, and $\theta(x)$ is a step function.  
The clock's Hamiltonian  is represented by $\py$,
and the time is recorded on the conjugate variable $\y$.
 \footnote{We have represented here the
ideal clock by a Hamiltonian
$H_{clock}=\py$ that is linear in the momentum.
This linear Hamiltonian can be obtained approximately for a 
free particle with $H={\bf P}_y^2/2M$. For a given duration $t$ 
we can approximate $H\simeq {\langle{\bf P}_y\rangle\over M}P_y+
const.$ 
by letting the mass be sufficiently 
large.  One could also consider a Larmor clock with a 
bounded Hamiltonian $H_{clock}=\omega {\bf J}$ \cite{peres}.  
The Hilbert space is
spanned by 2j+1 vectors where j is a natural number, and the
clock's resolution can be made arbitrarily fine by increasing j.}

The equations of motion read:
\beq
\dot \x =\px/m, \ \ \ \ \ \dot {\bf P}_\x = -\py\delta(\x)
\label{eq1}
\eeq
\beq
\dot \y = \theta(\x), \ \ \ \ \ \dot{\bf P}_\y=0.
\eeq
At $t\to \infty$ the clock shows the time of arrival:
\beq
\y_\infty = \y(t_0) + \int^\infty_{t_0} \theta(-\x(t)) dt
\eeq

A crucial difference between the classical and the quantum case, 
can be noted from Equation (\ref{eq1}). In the classical
case the back-reaction can be made negligible small  by choosing $P_y\to 0$.
In this case, the particle follows 
the undisturbed solution, $x(t)=x(t_0) + {p_x\over m}
(t-t_0)$. If initial
we set 
$y(t_0) = t_0$ and $x(t_0) <0$  the clock 
finally reads:
\beq
y_\infty =  y(t_0)+ \int^\infty_{t_0} 
\theta[-x(t_0)-{p_x\over m}(t-t_0)] dt
= -{m x(t_0)\over p_x}.
\eeq
The classical time-of-arrival is  $ t_A= y_\infty = -mx(t_0)/p_x$.
The same result would have been obtained by 
measuring the classical variable $-mx_0/p_x= -mx(t)/p_x +(t-t_0)$, 
at arbitrary time $t$. 
Consequently, the continuous and the indirect measurements
alluded to in Section I, are classically equivalent.
 
On the other hand, in quantum mechanics 
the uncertainty relation dictates a strong back-reaction, 
i.e. in the limit of $\Delta y=\Delta t_A \to 0$, 
$p_y$ in (\ref{eq1}) must have a large uncertainty, and the state 
of the particle must be strongly affected by the act of measuring.
Therefore, the two classically equivalent measurements become 
inequivalent in quantum mechanics.

Before we proceed to examine the continuous 
measurement process in more detail, we note that a more symmetric 
formulation of the 
above measurement exists in which knowledge of the 
direction from which particles are arriving is not needed. We can consider 
\beq
H = {1\over 2m} {\px}^2 + \theta(-\x) {\py}_1 +
\theta(\x) {\py}_2 .
\label{model}
\eeq
As before,  the particle's motion is confined to one 
spatial dimension, $x$.
Two clocks are represented by $\py_1$ and $\py_2$, 
and time is recorded on the conjugate variables $\y_1$ and
$\y_2$,  respectively.

The first clock operates only when the particle is located at  
$x<0$ and the second clock at $x>0$.
For example, if we start with a beam of particle at $x<0$, 
a measurement at $t\to\infty$ of $\y_1$ 
gives the time-of-arrival. 
Alternatively we could measure $t-\y_2$.
As a check we have $\y_1+\y_2=t$.
It is harder to determine the time-of-arrival if the 
particle arrives from both directions. If however it is 
known that initially 
$|x|< L$, we can measure $\y_1$ and $\y_2$ 
after  $t>>L/v$.
The time-of-arrival
will then be given by $t_A=\min(\y_1,\y_2)$.

Let us examine this system in more detail. For simplicity
we shall consider the case of only one clock and a particle 
initially at $x<0$, which travels towards the clock at $x=0$.
The eigenstates of the Hamiltonian are
\beq
\phi_{kp}(x,y,t) = \left\{ \begin{array}{ll}
(e^{ikx}+A_Re^{-ikx})e^{ipy-i\omega( t)} & x < 0 \\
A_Te^{iqx+ipy-i\omega(t)} & x \geq 0 
\end{array}
\right. 
\eeq
where  $k$ and $p$, are the momentum of the 
particle and the clock, respectively, and 
$\omega(t)=\frac{k^2t}{2m}+pt$. 
Continuity of $\phi_{kp}$ requires that 
\beqa
A_T & = & \frac{2k}{k+q} \nonumber\\
A_R & = & \frac{k-q}{k+q} ,
\eeqa  
where $q=\sqrt{k^2 + 2mp} =\sqrt{2m(E_k +p)}$.

The solution of the Schr\"odinger equation is 
\beq
\psi(x,y,t)= N\int^\infty_{-\infty}dk \int_0^\infty 
dp f(p) g(k) \phi_{kp}(x,y,t) , \label{eq:psi}
\eeq
where $N$ is a normalization constant and $f(p)$ and 
$g(k)$ are some  distributions.  
For example, with 
\beqa
f(p) & = & e^{-{\Delta_y}^2 (p-p_0)^2} \nonumber\\
g(k) & = & e^{-{\Delta_x}^2 (k-k_o)^2 + ikx_0}. \label{eq:distributions}
\eeqa
and $x_0>0$, the particle is initially localized
on the left ($x<0$) and the clock 
(with probability close to 1) runs.
The normalization in eq. (\ref{eq:psi}) is thus
$N^2=\frac{\Delta x \Delta y}{2 \pi^3}$.
By choosing $p_0 \approx 1/\Delta_y$, we can 
now set the the clock's
energy in the range $ 0 < p < 2/\Delta_y$.  

Let us first show that in the stationary point approximation 
the clock's final wave function is indeed centered around the 
classical time-of-arrival. 
Thus we assume that $\Delta_y$ and $\Delta_x$ are 
large such that $f(p)$ and $g(k)$ are sufficiently peaked. 
For $x>0$, the integrand in (\ref{eq:psi}) 
has an imaginary phase
\beq
\theta=qx+kx_o+py-\frac{k^2t}{2m}-pt .
\eeq
$\frac{d\theta}{dk}=0$ implies
\beq
x_{peak}(p) = -\frac{q(k_0)}{k_0}x_o+\frac{q(k_0)t}{m} ,
\eeq
and $\frac{d\theta}{dp}=0$ gives
\beq
y_{peak}(k)   =  t-\frac{mx}{q_0} . \nonumber\\ 
\eeq
Hence at $x=x_{peak}$ the clock coordinate $y$ is 
peaked at the classical time-of-arrival
\beq
y  =  \frac{mx_o}{k_0} .
\eeq

To see that the clock yields a reasonable record of the 
time-of-arrival, let us consider further the probability 
distribution of the clock
\beq
\rho(y,y)_{x>0} = \int dx |\psi(x>0,y,t)|^2 .
\eeq
In  the case of inaccurate measurements with a small back-reaction on the particle 
$A_T\simeq 1$. The  clocks density matrix is then found
(see Appendix) to be  given by:
\beq
\rho(y,y)_{>0} \simeq  
 \frac{1}{ \sqrt{2\pi\gamma(y)}} 
e^{-\frac{(y- t_c)^2}{2\gamma(y)}}  
%\label{eq:clockdist}
\eeq  
where the width is $\gamma(y)=\Delta y^2 + (\frac{m \Delta x}{k_o})^2
 + (\frac{y}{2 k_o \Delta x  })^2 $.  
As expected, the distribution is centered around the classical time-of-arrival $t_c=x_o m/k_o$.  The spread in $y$ has a term due 
to the initial width $\Delta y$ in clock position y.  The second and third term in $\gamma(y)$ is due to the kinematic spread in the time-of-arrival $\frac{1}{dE} =
\frac{m}{k dk}$ and is given by $ \frac{dx(y)\,m}{k_o}$ 
where $dx(y)^2=\Delta x^2 + (\frac{y}{2m\Delta x})^2$. The y dependence in the width in $x$ arises because the wavefunction is spreading as time increases, so that at later $y$, the wavepacket is wider.  As a result, 
the distribution differs slightly from a Gaussian although this effect is 
suppressed for particles with larger mass.

When the back-reaction causes a small disturbance to the particle, the 
clock records the time-of-arrival. What happens when we wish to make more accurate measurements?  
Consider the exact transition probability  $T=\frac{q}{k}|A_T|^2$,
which also determines
the probability to stop the clock. The latter is given by
\beq
\sqrt{\frac{E_k+p}{E_k}}
\Biggl[ {2\sqrt{E_k}\over \sqrt{E_k} + \sqrt{E_k+p} }\Biggr]^2 .
\eeq
Since the possible values obtained by $p$ are of the order
$1/\Delta_y  \equiv  1/\Delta t_A$, 
the probability to trigger the clock remains of order one
only if  
\beq
\bar E_k \Delta t_A > 1 .
\label{Edt}
\eeq
Here $\Delta t_A$ stands for the initial uncertainty 
in position of the dial $\y$ of the clock,
and is interpreted as the accuracy of the clock.
$\bar E_k$ can be taken as the typical initial kinetic energy of 
the particle.

In measurements with accuracy better then $1/\bar E_k$
the probability to succeed drops to zero
like $\sqrt{E_k \Delta t_A}$, and the time-of-arrival
of most of the particles cannot be detected.
Furthermore,  the probability distribution 
of the fraction which has been  detected depends on the accuracy 
$\Delta t_A$ and can become distorted with increased accuracy.
This observation becomes apparent in the following simple example.
Consider an  initial  wave packet that is 
composed of a superposition of two Gaussians centered around $k=k_1$
 and 
$k=k_2>>k_1$. Let the classical time-of-arrival of the two Gaussians
be $t_1$ and $t_2$ respectively. When the inequality (\ref{Edt})
is satisfied, two peaks around $t_1$ and $t_2$ will show up in 
the final probability distribution. 
On the other hand, for ${2m\over k_1^2}>
\Delta t_A > {2m\over k_2^2}$, 
 the time-of-arrival of the less energetic peak will 
contribute  less to the distribution 
in $y$, because it is less likely to trigger the
clock.  Thus, the peak at $t_1$ will be suppressed.
Clearly,  when the precision is finer than $1/\bar E_k$
we shall obtain a distribution which is 
considerably different from that obtained for the case 
$\Delta t_A > 1/\bar E_k$ when the two peaks contribute equally.

\subsection{Two-level detector with a clock}

A more realistic set-up for a time-of-arrival measurement
is one that also includes a particle detector
which switches the clock off as the particle arrives.
We shall describe the particle detector as a two-level
spin degree of freedom. 
The particle will flip the state of the trigger from 
"on" to "off", ie. from $\up_z$ to $\down_z$.
First let us consider a model for the trigger without
including the clock: 
\beq
H_{trigger} = {1\over 2m} {\bf P_x^2} + {\alpha\over2}(1 +
\sigma_x) \delta({\bf x}) .
\eeq
The particle interacts with the repulsive
Dirac delta function potential at $x=0$, only if the spin is in
the 
$|\up_x\ra$ state, or with a vanishing potential if the state is 
$|\down_x\ra$. In the limit 
$\alpha \to \infty$ the potential becomes totally reflective
(Alternatively, one could  have considered 
a barrier of height $\alpha^2$ and  width $1/\alpha$.)
In this limit, consider a state of an incoming particle and 
the trigger in the "on" state: $|\psi\ra |\up_z\ra$.
This state  evolves to 
\beq
|\psi\ra |\up_z\ra  \ \ \to \ \ 
 {1\over \sqrt2}
 \bigg[ |\psi_R\ra|\up_x\ra  + |\psi_T\ra  |\down_x\ra \biggr] ,
\eeq
where $\psi_R$ and $\psi_T$ are the reflected and transmitted 
wave functions of the particle, respectively.

The latter equation can be rewritten as 
\beq
{1\over 2}|\up_z\ra (|\psi_R\ra +|\psi_T\ra)
+{1\over 2}|\down_z\ra(|\psi_R\ra - |\psi_T\ra)
\eeq

Since $\up_z$ denotes the "on" state of the trigger, and
$\down_z$
denotes the "off" state, we have flipped
the trigger from the "on" state to the "off" state with
probability $1/2$.
By increasing the number of detectors, 
this probability can be made as close as we like to one.
To see this, consider  $N$ spins as $N$ triggers 
and set the  Hamiltonian to be 
\beq
{\px}^2/2m + (\alpha/2)\Pi_n (1+\sigma_x^{(n)})\delta(\x) . 
\eeq
We will say that the particle has been detected if at least 
one of the spin has flipped.
One can verify that in this case 
the probability that at least one spin has flipped is now 
$1-2^{-N}$.

So far we have succeeded in recording the event of arrival to 
a point. We have no information at all on the time-of-arrival.
It is also worth noting that the net energy 
exchange between the trigger and the particle is zero, ie. the 
particle's energy is unchanged.

This model leads us to reject the arguments of Allcock.   He considers
a detector which is represented by a pure imaginary absorber $H_{int}=iV\theta(-\x)$. 
Allcock's claim is that measuring the time-of-arrival is equivalent to absorbing
a particle in a finite region. If you can absorb the particle in an arbitrarily short time, then you have succeeded in transferring the particle from an incident channel into a detector channel and the time-of-arrival can then be recorded. Using his interaction Hamiltonian one finds
that the particle is absorbed in a rate proportional to $V^{-1}$.  One can 
increases the rate of absorption by increasing $V$, but the particle will be reflected unless $V<<E_k$.  He therefore claims that since you cannot absorb the particle in an arbitrarily short time, you cannot record the time-of-arrival
with arbitrary accuracy.

However, our two level detector is equivalent to a detector which absorbs
a particle in an arbitrarily short period of time, and then transfers the information to another channel.  The particle is instantaneously
converted from one kind of particle (spin up), to another kind of particle
(spin down).  A model for arbitrarily fast absorption is also given in \cite{muga}, although in this case, the absorber does not work for arbitrary wavefunctions.  We therefore see that considerations of absorption alone do not place any restrictions on measuring the time-of-arrival.

However, we shall see that when we proceed to couple the trigger to a clock
we do find a limitation on the time-of-arrival. The total 
Hamiltonian is now given by
\beq
H_{trigger+clock} = 
{1\over 2m} {\bf P_x^2} + {\alpha\over2}(1 + \sigma_x)
\delta({\bf x}) +  {1\over2}(1 + \sigma_z){\bf P_y} .
\eeq
 
Since we can have $\alpha>> P_y$ it would seem that 
the triggering mechanism need not be affected by the clock.
If the final wave function includes a non-vanishing  amplitude
of  $\down_z$, the clock will be  turned off
and the time-of-arrival recorded. 
However, the exact solution shows that this is not the case.
Consider for example an initial state
of an incoming wave from the left and the spin in the $\up_z$
state.

The eigenstates of the Hamiltonian in the basis of $\sigma_z$
are
\beq
\Psi_L(x) = \left( 
\matrix{
e^{ik_\up x}+\phi_{L\up}e^{-ik_\up x} \cr 
\phi_{L\down} e^{-ik_\down x}  \cr }
\right) e^{ipy} ,
\eeq
for $x<0$ and 
\beq
\Psi_R(x) = \left( 
\matrix{
\phi_{R\up} e^{ik_\up x} \cr 
\phi_{R\down} e^{ik_\down x}  \cr }
\right) e^{ipy} ,
\eeq
for $x>0$.
Here $k_\up = \sqrt{2m(E-p)}=\sqrt{2mE_k}$
and $k_\down =\sqrt {2m E} = \sqrt {2m(E_k+p})$.
 
Matching conditions at $x=0$ yields 
\beq
\phi_{R\up} = {{2k_\up\over m\alpha} - {k_\up\over k_\down}
\over {2k_\up\over m\alpha} - ( 1 +{k_\up\over k_\down}) }
\eeq
\beq
\phi_{R\down} =  {k_\up\over k_\down}((\phi_{R\up}-1)=
{  {k_\up\over k_\down}
\over {2k_\up\over m\alpha} - ( 1 +{k_\up\over k_\down}) } ,
\label{phidown}
\eeq
and 
\beq
\phi_{L\down} = \phi_{R\down} 
\label{leftdown}
\eeq
\beq
\phi_{L\up} = \phi_{R\up} - 1 .
\label{leftup}
\eeq

We find  that in the limit $\alpha \to \infty$ the transmitted 
amplitude is 
\beq
\phi_{R\down} = - \phi_{R\up} = {\sqrt{E_k} \over
\sqrt{E_k} + \sqrt{E_k+p} } .
\eeq
Precisely as in the previous section, the transition 
probability decays like $\sqrt{E_k/p}$.
From  eqs. (\ref{leftdown},\ref{leftup}) we get that
$\phi_{L\down}\to 0$, and $\phi_{L\up} \to 1$ as the accuracy of
the clock increases. Hence the 
particle is mostly reflected back and the spin remains in the 
$\up_z$ state; i.e., the clock remains in the "on" state.  

The present model gives rise to the same difficulty as the
previous model. Without the clock, we can flip 
the "trigger" spin by means of a localized interaction, but when
we couple the particle to the clock, the probability 
to flip the  spin and turn the clock off decreases
gradually to zero when the clock's precision is improved.

\subsection{Local amplification of Kinetic Energy}

The difficultly with the previous examples seems to be that
the particle's kinetic energy is not sufficiently large, and 
energy can not be exchanged with the clock.
To overcome this difficulty one can imagine  introducing  a 
``pre-booster'' device just before the particle arrives 
at the clock. If it could   
boost the particle's kinetic  energy arbitrarily high, 
without distorting the incoming 
probability distribution (i.e. amplifying all wave components $k$
with the same probability),  
and at an arbitrary short distance from the clock, 
then the time-of-arrival could be measured to arbitrary accuracy.
Thus, an equivalent problem is: can we  
boost the energy of a  particle by 
using only localized (time independent) interactions?
  
Let us consider the following toy model of an energy booster
described by the Hamiltonian
\beq
H={1\over 2m} {\bf P_x^2} +\alpha\sigma_x\delta(\x) +\frac{W}{2}\theta(\x)(1+\sigma_z)
+\frac{1}{2}[V_1\theta(-\x)-V_2\theta(\x)](1-\sigma_z). 
\label{booster}
\eeq
Here, $\alpha, W, \ V_1$ and $V_2$ are positive constants.
Let us consider an incoming wave packet propagating from left to right.
The role of the term $\alpha \sigma_x \delta(x)$ is to flip the spin $\up_z$
to $\down_z$. The $\up_z$ component of the wave function is damped out 
exponential by the $W$ term for  $x>0$. The 
 $\down_z$ component is damped out for  $x<0$ by the term
$V_1$, but increases its  kinetic energy for  $x>0$  by $V_2$.  
As we shall see, for a given momentum $k$, one can chose the four 
free parameters above such that 
the wave is transmitted through the booster with probability $1$, 
while the gain in energy $V_2$ can be made
arbitrarily large. On  the other hand, 
the potential barrier $W$  can be made arbitrarily large. 
The last requirement means  that the unflipped component, decays for 
$x>0$ on arbitrary short scales, which allows us to locate the booster 
arbitrarily close to the clock, while preventing destructive interference 
between the flipped and un-flipped transmitted waves.

The eigenstates of (\ref{booster}), in the basis of $\sigma_z$,
are given by
\beq
\Psi_L(x) = \left( 
\matrix{
e^{ik x}+\phi_{L\up}e^{-ik x} \cr 
\phi_{L\down} e^{q x}  \cr }
\right)  
\eeq
for $x<0$ and 
\beq
\Psi_R(x) = \left( 
\matrix{
\phi_{R\up} e^{-\lambda x} \cr 
\phi_{R\down} e^{ik' x}  \cr }
\right) 
\eeq
for $x>0$,
where $k^2=V_1-q^2=-\lambda^2+W=-V_2+k'^2$. 
Matching conditions at $x=0$ we find
\beq
\phi_{L\up}= \phi_{R\up}-1 = {k'k+q \lambda  +i(kq-k'\lambda)-\alpha^2 \over
k'k -q\lambda +i (k'\lambda +kq) + \alpha^2}, 
\eeq
\beq
\phi_{R\down}=\phi_{L\down} = {\alpha\over ik'-q} \Bigl( 1+ \phi_{L\up}
\Bigr).
\eeq 
For a given $k, W$ and $V_2$ (or given $k,\ \lambda$ and $ k'$)
we still are free to chose $\alpha$ and $V_1$ (or $q$).
We now demand that 
\beq
\alpha = k'k + q\lambda , \ \ \ \ q= \lambda {k'\over k}.
\eeq
With this choice we obtain:
\beq
J_{L\up} = 0 , \ \ \ \  J_{R\down} = 
{k'\over k}|\phi_{R\down}|^2 = 1.
\eeq
Therefore, the wave has been fully transmitted and the spin has flipped 
with probability $1$. 

So far we have considered an incoming wave with fixed momentum $k$.
For a general incoming wave packet only a part of the wave will 
be transmitted and amplified. Furthermore one can verify that the 
amplified transmitted wave has a 
different form than the original wave
function since different momenta have been amplified with different
probabilities.
Thus, in general, although amplification is possible and
indeed will lead to a much higher rate of detection,
it will give rise to a distorted 
probability distribution for the time-of-arrival.  

There is however one limiting case in which the method does seem to 
succeed.
Consider a narrow wave peaked around $k$ with a width $\delta k$.
To first order in $\delta k$, the probability $T_\down$ that the particle
is successfully boosted is given by
\beq
J_{R\down}\simeq 1+
\frac{2\delta k}{k}. 
\eeq 
Therefore in the special case that ${dk\over k} <<1$, the transmition 
probability is still close to one. If in this case we known 
in advance the value of $k$ up to $\Delta k << k$, we can indeed use the 
booster to improve the bound (\ref{Edt}) on the accuracy.

The reason why this seems to work in this limiting case is as follows.
The probability of flipping the particle's spin 
depends on how long it spends in the magnetic field described  by 
the $\alpha$ term in (\ref{booster}).  
If however, we know beforehand,
how long the particle will be in this field, then we can tune the
strength of the magnetic field ($\alpha$) 
so that the spin gets flipped.  
The requirement that 
$ \Delta k  / k   <<1 $ is thus equivalent to having a small uncertainty in 
the ``interaction time'' with this field. 
It must be emphasized however, that these measurements
cannot be used for general wave functions, and that even in the special case above, 
one still requires some prior 
information of the incoming wave function.

\subsection{Gradual triggering of the clock}

In order to avoid the reflection found in the previous two 
models, we shall now replace the sharp 
step-function interaction between the clock and particle by a
more gradual transition. 

When the WKB condition is satisfied 
\beq
{d\lambda(x)\over dx} = \epsilon << 1
\label{wkb}
\eeq
where $\lambda(x)^{-2} = 2m[E_0-V(x)]$, the reflection amplitude 
vanishes as 
\beq
\sim \exp (-1/\epsilon^2)
\eeq
Solving the equation for the potential with 
a given $\epsilon$
we obtain
\beq
V_\epsilon(x) = E_0 - {1\over 2m\epsilon^2} {1\over x^2}
\eeq
Now we  observe that any particle with 
$E\ge E_0$ also satisfies the WKB condition (\ref{wkb}) above
for the $same$ potential $V_\epsilon$.
Furthermore $p_y V_\epsilon$ also satisfies the condition for 
any $p_y > 1$.

These considerations suggest that we should replace the 
Hamiltonian in eq. (\ref{model}) with 
\beq
H = {\px^2/2m} + V(x)\py
\label{conti}
\eeq
where 
\beq
V(x) = \left\{ \begin{array}{ll}
-{x_A^2\over x^2} & x < x_A \\
-1  & x \ge x_A
\end{array}
\right. 
\eeq
Here $x_A^{-2} = 2m\epsilon^2$.

Thus this model describes a gradual triggering $on$ of the clock
which takes place when the particles propagates from $x\to
-\infty$
towards $x=x_A$. In this case the arrival time is 
approximately given by $t- \y$ where $t= t_{f} - t_{i}$.
Since without limiting the accuracy of the clock
we can demand that $p_y >> 1$, the 
reflection amplitude off the potential step is exponentially 
small for $any$ initial 
kinetic energy $E_k$.

The problem is however that the final value of $t-\y$ does not 
always correspond to the time-of-arrival since 
it contains errors due to 
the affect of the potential $V(x)$ on the particle which we 
shall now proceed to examine.

In the following we shall ignore ordering problems and 
solve for the classical equations of 
motion for (\ref{conti}).  
We have
\beq
y(t_f) - y(t_i) = \int_{t_i}^{t_f} V(\x(t')) dt'
\eeq
which can be decomposed to 
\beq
y(t_f) - y(t_i) =  (t_i- t_0) + (t_f-t_i)
+\int_{t_i}^{t_0}V(x(t'))dt' 
\equiv A + B + C 
\label{composite}
\eeq
where
\beq
A=  {1\over\sqrt{2mE}} \biggl[ 
\sqrt{x_A^2 + p_y  x^2_A/E} - \sqrt{x_i^2 + p_y x_A^2/E} 
\biggr]
\eeq
is the time that the particle travels from $x_i$ to  $x_A$
in the potential $p_y V(x)$,
$B$ is the total time, and 
\beq
C = -{x_A\over\sqrt{2mp_y}}\biggl[
\log{ 1+ \sqrt{1 + {E\over p_y}}\over 1+
 \sqrt{1 + {Ex_i^2\over p_yx_A^2}}  } + \log{x_i\over x_A}
\biggr]
\eeq
The last term $C$, corresponds to an error due to the 
imperfection of the clock, i.e. the motion of the clock
prior to arrival to $x_A$. By making 
$p_y$ large we can minimize the error from this term to
$ \sim  (x_A \log p_y /\sqrt {2mp_y} )$ .
 
Inspecting equation (\ref{composite}) we see that by 
measuring $y_f-y_i$ and then subtracting $B=t_f-t_i$
(which is measured by another clock)  
we can determine
the time $t_0-t_i$, which is the time-of-arrival 
for a particle in a potential $p_y V(x)$, up to the 
correction $C$. However this time
reflects the motion in the presence of  an external 
(unknown) potential, while we 
are interested in 
the time-of-arrival for a free particle.

Nevertheless, if $p_y/E << 1$ 
we obtain
\beq
-A = {x_A - x_i\over \sqrt{2mE} } + O\biggl({p_y\over E}\biggr)
\eeq
The time-of-arrival can hence be measured provided 
that  $E_k \Delta t >> 1$. 
On the other hand, when the detector's accuracy is  
$\Delta t< 1/E$, the particle
still triggers the clock. 
However the measured quantity, $A$, no longer 
correspond to the time-of-arrival. 
Again, we see that 
when we ask for too much accuracy,
the particle is strongly disturbed and the 
result has nothing to do with the time-of-arrival of 
a free particle.

\subsection{General considerations}

We have examined several  models for a measurement of time-of-arrival and found 
a limitation, 
\beq
\Delta t_A > 1/\bar E_k ,
\label{limit}
\eeq
on the accuracy that $t_A$ can be measured. 
Is this limitation a general feature of quantum mechanics?

First we should notice that eq. (\ref{limit}) does not seem to
follow
from the uncertainty principle. 
Unlike the uncertainty principle, whose origin is kinematic, 
(\ref{limit}) follows from the nature of the  
$dynamic$ evolution of the system during a measurement.
Furthermore here we are considering a restriction on the measurement of a 
single quantity. 
While it is difficult to provide a general proof,  
in the following we shall indicate why 
(\ref{limit}) is expected to  hold
under more general circumstances.

Let us examine the basic features that  
gave rise to (\ref{limit}). 
In the toy models considered in sections IIA and  IIB, 
 the clock and the particle
had to exchange energy $p_y \sim 1/\Delta t_A$.
The final kinetic energy of the particle is larger by $p_y$.
As a result, the effective interaction by which the clock switches off, 
looks from the point of view of the particle like a 
step function potential. This led to  
``non-detection'' when (\ref{limit}) was violated.

Can we avoid this energy exchange between the 
particle and the clock? 
Let us try to deliver
this energy to some other system without modifying 
the energy of the particle. 
For example consider the following Hamiltonian for 
a clock with a reservoir:
\beq
H = {\px^2\over 2m} + \theta(-\x)H_c + H_{res} + V_{res}\theta(\x)
\eeq
The idea is that when the clock stops, it dumps 
its energy into the 
reservoir, which may include many other degrees of freedom, 
 instead of delivering it to the particle.
In this model, the particle is coupled directly 
to the clock and reservoir,
however we could as well use the 
idea of section IIB above.
In this case: 
\beq
H = {\px^2\over 2m} +{\alpha\over2}(1 + \sigma_x)
\delta({\bf x}) +  {1\over2}(1 + \sigma_z){H_c}+ 
 H_{res} + {1\over2}(1 - \sigma_z)V_{res} .
\eeq
The particle detector has the role of providing a coupling
between  the clock and reservoir.
 
Now we notice that in order to transfer the clock's
energy to the reservoir without affecting the free particle,  
we must also prepare the clock and reservoir 
in an initial state that satisfies the condition  
\beq
H_c - V_{res} =0
\eeq
However this condition does not commute with the clock time 
variable $\y$.  We can 
measure initially $\y - {\bf R}$, where $R$ is a collective 
degree of freedom of the reservoir such that $[{\bf R}, V_{res}]
= i$, 
but in this case we shall not gain information 
on the time-of-arrival $y$ since $R$ is unknown.
We therefore see that in the case
of a sharp transition, i.e. for a localized interaction with the 
particle, one cannot avoid a shift in the particle's energy.
The "non-triggering" (or reflection)
effect cannot be avoided.

We have also seen that the idea of boosting the particle ``just before''
it reaches the detector, fails in the general case. What happens in this case
is that while the detection rate increase, one generally destroys the 
initial information stored in the incoming wave packet. 
Thus although higher accuracy measurements are now possible, they do not 
reflect directly the time-of-arrival of the initial wave packet.

Finally we note that in reality, measurements usually 
involve some type of cascade 
effect, which lead to signal amplification and finally allows 
a macroscopic clock to be triggered. 
A typical example of this type would be the photo-multiplier where 
an initially small energy is amplified gradually and 
finally detected.  
Precisely this type of process occurs also in 
the model of section IID.
In this case the particle gains energy  
gradually by ``rolling down'' a smooth step function.
It hence always triggers the clock.
The basic problem with such a detector is that 
when (\ref{limit}) is violated, the ``back reaction''
of the detector on the particle, 
during the gradual detection, becomes large.
The relation between the final record to  
the quantity we wanted to measure is lost.

\section {Conditions on A Time-of-Arrival operator}

As discussed in the introduction, although a direct measurement
of the time-of-arrival may not be possible, one can still try 
to observe it indirectly by measuring 
some operator $\T(\p, \x, x_A)$.
In the next two  sections  we shall  examine this 
operator and its relation to the continuous measurements described 
in the previous sections. 
First in this section we show that an exact time-of-arrival operator
cannot exist for systems with bounded Hamiltonian.

To begin with, let us start with the assumption that the 
time-of-arrival is described, as other observables in quantum 
mechanics, by a Hermitian operator $\T$. 
\beq
\T(t) |t_A\ra_t = t_A |t_A\ra_t
\label{eigen}
\eeq
Here the subscript $\ra_t$ denotes the time dependence of 
the eigenkets, and 
$\T$ may depend explicitly on time.
Hence for example, the probability distribution 
for the time-of-arrival for the state
\beq
|\psi\ra = \int g(t_A') |t_A'\ra dt_A'
\eeq
will be given by  $prob(t_A) = |g(t_A)|^2$. 
We shall now also assume that the spectrum of 
$\T$ is continuous and 
unbounded: $-\infty<t_A<\infty$.

Should $\T$ correspond to time-of-arrival it 
must satisfy the following obvious condition.
${\T}$ must be a  constant of motion and in the Heisenberg representation
\beq 
 {d{\T}\over d t}= 
 {\partial{\T}\over \partial t} +
{1\over i} [{\T}, H] = 0. 
\label{constant}
\eeq
That is, the time-of-arrival cannot change in time.

For a time-independent Hamiltonian, time translation invariance
implies that the eigenkets  $|t_A\rangle_t$ depends only on $t-t_A$, 
i.e. the eigenkets  cannot depend on the absolute time $t$. 
This means for example that at the time of arrival: $|t_A\rangle_{t=t_A}=
|t_A'\rangle_{t=t_A'}$.
Time-translation invariance implies
\footnote{In Allcock's proof \cite{allcock1} of the non-existence of a time-of-arrival 
operator for the special case of a free Hamiltonian, he assumes that
$|t_a + \tau\rangle = e^{-i\tau H} |t_a\rangle$. However, by definition, a time-of-arrival eigenstate which will arrive at the time $t_a$ will remain an
eigenstate which arrives at $t_a$ as the system evolves.  Allcock's proof is 
thus a proof 
of the non-existence of a time operator -- not time-of-arrival operators.}
\beq
|t_A\rangle_t = e^{-i\G} |0\rangle_0 .
\eeq
where $\G= \G(t-t_A)$ is a hermitian operator.
Therefore,  $|t_A\rangle_t$ satisfies the differential equations
\beq
i {\partial\over \partial t_A} |t_A\rangle_t  = {\partial \G\over 
\partial t_A} |t_A\rangle_t= -{\partial \G\over 
\partial t} |t_A\rangle_t, \ \ \ \ 
i {\partial\over \partial t} |t_A\rangle_t  = {\partial \G\over 
\partial t} |t_A\rangle_t .
\eeq
Now act on  the eigenstate equation (\ref{eigen}) with the differential operators
$i\partial_{t_A}$ and $i\partial_t$. This yields
\beq
-\T {\partial \G\over  \partial t} |t_A\rangle_t = - t_A {\partial \G\over 
\partial t} |t_A\rangle_t + i|t_A\rangle_t , 
\eeq 
and 
\beq
i{\partial \T\over \partial t} |t_A\rangle_t  + \T {\partial \G\over 
\partial t} |t_A\rangle_t = 
t_A {\partial \G\over 
\partial t} |t_A\rangle_t . 
\eeq
By adding the two equations above, the dependence on $\partial \G/\partial t$ 
drops off,  and after using 
the constancy of $\T$ (eq. \ref{constant}) 
we get  
\beq
\biggl( [\T,H] + i\biggr)|t_A\rangle = 0. 
\eeq
Since  the eigenkets $|t_A\rangle$ span,  by assumption,  the full Hilbert 
space   
\beq
[\T, H] = -i . 
\label{conj}
\eeq
Hence $\T$ is a generator of energy translations. From equation (\ref{constant})
we have $\T= t - \hat {\bf T}$, where $\hat {\bf T}$ is the  
``time operator'' of the system whose Hamiltonian 
is $H$.  It is well know that equation (\ref{conj}) 
is inconsistent unless the Hamiltonian is unbounded from above 
and below \cite{pauli}.
\section{Measuring the Time-of-Arrival Operator vs. Continuous Measurements}
Although formally there cannot exist a time-of-arrival operator $\T$, it
may be possible to approximate $\T$ to arbitrary accuracy 
\cite{rovelli}.  Kinematically, one expects that the time-of-arrival operator for a free particle arriving at the location $x_A=0$ might be given by 
\beq
\T=-\frac{m}{2}\frac{1}{\sqrt{\p}}\x (0)\frac{1}{\sqrt{\p }}. 
\eeq
The choice for the time operator is clearly not unique. An equally 
valid choice is $-m({1\over \p}\x +\x {1\over \p})$, etc.
Furthermore, since $\T$  is ill defined at p=0, it's eigenvalues 
\beq
\langle k | T^\pm \rangle = \theta(\pm k)\sqrt{\frac{k}{2\pi m}}
e^{i\frac{Tk^2}{2m}}   
\eeq
are not orthogonal.  
\beq
\langle T | T' \rangle = \delta(T-T') - \frac{i}{\pi(T-T')} .
\eeq
Thus, $\T$ is not Hermitian.
We can however, define the regularized
Hermitian operator $\T ' = OTO$ where $O= 1 - |p=0><p=0|$.  
It's eigenvalues are
complete and orthogonal, and it circumvents the proof given above, 
because it satisfies  $ [\T ', H] = i\hbar O$
i.e. it is not conjugate to $H$ at $p=0$.  Although $\T$ is not
always the shift operator of the energy, the measurement can be carried
out in such a way that this will not be of consequence.  To see this,
consider the interaction Hamiltonian
\beq
H_{meas} = \delta(t) {\bf q} \T ', 
\eeq
which modifies the initial wave function $\psi \to \exp(-iqT')\psi$.
We need to demand that $\T '$ acts as a shifts operator of the energy 
of $\psi$ during the measurement.  Therefore we need that 
$q>-E_{min}$,  where $E_{min}$ is the minimal energy
in the energy distribution of $\psi$. In this way, the measurement 
does not shift the energy down to $E=0$ where $\T '$ 
is no longer conjugate to $H$.  The value of $\T '$ is recorded on 
the conjugate of $q$ -- call it $P_q$.  Now the uncertainty is given by
$dT_A' = d(P_q) = 1/dq$, thus naively from $dq=1/dT_A' < E_{min}$,  
we get $E_{min} dT'>1$.  However here, the average $ \la q \ra$ was 
taken to be zero.  There is no reason not to take $\la q \ra$ to be much 
larger than $E_{min}$,  so that $\la q \ra -dq >> -E_{min}$.  If we do so, 
the measurement increases the energy of $\psi$ and $\T '$ is always 
conjugate to $H$.  The limitation on the accuracy is in this case
$dT_A' > 1/ \la q \ra$ which can be made as small as we like.\\
Nevertheless, there are still problems with this time-of-arrival operator.
One finds that at the time of arrival, the eigenstates of $\T$,
$\langle x | T(t_A)\rangle$ are not delta functions $\delta(x)$ but 
are proportional to $x^{-3/2}$.  This means that at the time of arrival,
the particle has a probability of not being found at the origin.
Furthermore, if ${\bf P}_{0}$ is the projector onto $x=0$, one finds that
\beq
\langle\psi [\T, {\bf P}_{0}]|\psi\rangle=-\frac{i}{2}Re\{\psi(x=0)\int dk
\psi^*(k)\frac{m}{k^2}\}. 
\eeq  
A measurement of the time-of-arrival operator is not equivalent to 
continuously monitoring the point-of-arrival.  Furthermore, if one
measures a time-of-arrival operator at a time $t'$ before the particle 
arrives, then one needs to know the full Hamiltonian from time $t'$ until
$t_A$.  Even if one knows the full Hamiltonian, and can find an approximate time-of-arrival operator, one has to have faith that the Hamiltonian will not 
be perturbed after the measurement has been made. On the other hand, the continuous measurements we have described can be used with any Hamiltonian. 

Finally, how does the resulting measurement of a time-of-arrival operator
compare with that of a continuous measurement? 
From the discussion in Section IIA, it should be clear that in the limit
of high precision, continuous measurements respond very differently in comparison
to the time operator. At the limit of $dt_A\to 0$ all the particles
bounce back from the detector. Such a behavior does not occur for 
the time of arrival operator. Nevertheless, 
one may still hope that since the eigenstates 
of $\T$ have an  infinitely spread in energy, they do trigger
a clock   even if $dt_A\to0$.
For the type of models we have been considering,  
we can show however that this will not happen.

Let us assume that the interaction of one eigenstate of $\T$
with the clock (of,  say,  Section IIA) evolves as
\beq
|t_A\rangle|y=t_0\rangle \ \rightarrow   \ 
|\chi(t_A)\rangle|y=t_A\rangle  + |\chi'(t_A)\rangle|y=t\rangle.
\label{map}
\eeq
Here, $|y=t_0\rangle$ denotes an initial state of the clock with $dt_A\to0$, 
$|\chi(t_A)\rangle$ denotes the final state of the particle if the clock 
has stopped, and $|\chi'(t_A)\rangle$ the final state of the particle 
if the clock has not stopped.

Since the eigenstates of $\T$ form a 
complete set,  we can express any state of the 
particle as $|\psi\rangle= \int dt_A C(t_A) |t_A\rangle$.
We then obtain :
\beq
\int dt_A C(t_A) |t_A\rangle |y=t_0\rangle  \ \rightarrow  \ 
\int dt_A C(t_A) |\chi(t_A)\rangle |y=t_A\rangle +  \biggl( \int dt_A C(t_A) 
 |\chi'(t_A)\rangle \biggr) |y=t\rangle .
\eeq
The final probability to measure the time-of-arrival 
is hence $\int dt_a |C(t_a)\chi(t_a)|^2$. On the other hand 
we found that for a general wave function $\psi$, 
in the limit of  $dt_a\to 0$, the probability 
for detection vanishes.
Since the states of the clock, $|y=t_a\rangle$, 
 are orthogonal in this limit, this implies that  $\chi(t_a) = 0 $
in eq. (\ref{map}) for all $t_A$.
Therefore, the eigenstates of $\T$ cannot trigger the clock.

\section{Conclusion}
We have examined various models for  
the measurement of time-of-arrival, $t_A$,
and found a basic limitation on 
the accuracy that $t_A$ can be determined reliably:
$\Delta t_A > 1/\bar E_k$.
This limitation is quit different in origin 
from that due to the 
uncertainty principle; here it applies to a $single$
quantity.
Furthermore, unlike the kinematic nature of the uncertainty
principle, in our case the limitation is essentially
dynamical in its origin; it arises when  the time-of-arrival is
measured 
by means of a continuous interaction between the 
measuring device and the particle. 

 We would also like to  stress, that continuous measurements,
differ both conceptually and quantitatively 
from a measurement  of the time-of-arrival operator.
Operationally one performs here two completely different measurements.
While the time-of-arrival operator is a formally
 constructed operator which can be measured by an impulsive
von-Neumann interaction, it seems
that continuous measurements are much more closer to actual
 experimental set-ups.
Furthermore, we have seen that the result of these two measurements 
do not need to agree,   in particular  
in the high accuracy limit, continuous measurements 
give rise to entirely different behavior.
This suggests that as in the case of the problem of finding 
a ``time operator'' \cite{unruh-wald} for closed quantum systems, 
the time-of-arrival operator has a somewhat limited physical meaning.

\vspace {2 cm} 

{\bf Acknowledgment}

B.R., W.G.U., and J.O. 
would like to thank Asher Peres for discussions.
Y. A. acknowledges the support of  the Basic
Research Foundation,  grant 614/95,
administered by the Israel Academy of Sciences and Humanities.

\vspace {3 cm}

\appendix
\section{}
Using the simple model of section IIA (\ref{model1clock}), we now calculate the probability distribution of a clock which measures the time-of-arrival of a Gaussian wave packet. 
We will perform the calculation in the limits when 
the clock is extremely accurate and extremely inaccurate.  The wave function
of the clock and particle is given by (\ref{eq:psi}) and the distributions are 
both Gaussians given by (\ref{eq:distributions}). In the
inaccurate limit, when $p_o<<k$, $A_T\sim 1$.
We trace over the position of the particle on the condition that the
clock was triggered, ie. $x>0$.
\beqa
\rho(y,y)_{x>0}&=&\int dx |\psi(x>0,y,t)|^2 \nonumber\\
&\simeq& N^2 \int_{-\infty}^\infty dk\, dk' 
\int_0^\infty dp\, dp'\, dx  g(k)g^*(k')f(p)f^*(p') 
e^{i(q-q')x+i(p-p')y-\frac{i(q^2-q'^2)t}{2m}} 
\eeqa
After a sufficiently long time, ie. $t>>t_a$ the wave function has no
support
on the negative x-axis, and if $p_o > 1 / \Delta y$, then it will not have 
support in negative $p$.  We can thus integrate $p$ and $x$ over the
entire axis.  Integrating over $x$ gives a delta-function in q.  We can then
integrate over $p'$ to give
\beqa
\rho(y,y)_{x>0} &\simeq& \frac{2 \pi N^2}{m} \int dk\,dk'\, dp 
\sqrt{k^2+2mp} \,  g(k)g^*(k')f(p)f^*(p+\frac{k^2-k'^2}{2m}) 
e^{i(k'^2-k^2)\frac{y}{2m}}
\eeqa
where we have used the fact that $\delta(f(z))=\frac{\delta(z-z_o)}{f'(z=z_o)}$ when $f(z_o)=0$.
The square root term varies little in comparison with 
the exponential terms and can be replaced by its average value $\sqrt{k_o^2+2mp_o}\simeq k_o$. 
% $f(p)f^*(p+\frac{k^2-k'^2}{2m})$.  This term is a Gaussian peaked around %$p_o-\frac{k^2+k'^2}{4m}$ and we can therefore 
%replace $p$ in the square root by the value where the exponential is peaked.
Integrating over $p$ gives
\beq
\rho(y,y)_{x>0} \simeq \frac{2 \pi N^2 k_o}{m}\sqrt{\frac{\pi}{2\Delta y^2}} 
\int dk\,dk'\,  e^{\frac{-\Delta y^2}{8m^2}(k+k')^2(k-k')^2}
g(k)g^*(k') e^{i(k'^2-k^2)\frac{y}{2m}}.
\eeq
Since $\Delta y\, k >> 1 $, for a wave packet peaked around $k_o$ we can
approximate the argument of the first exponential by $\frac{-\Delta y^2k_o^2}{2m^2}(k-k')^2$.  This allows us to integrate over $k$ and $k'$
\beq
\rho(y,y)_{>0} \simeq  
 \frac{1}{ \sqrt{2\pi\gamma(y)}} 
e^{-\frac{(y- t_c)^2}{2\gamma(y)}}  
\label{eq:clockdist}
\eeq  
where the width is $\gamma(y)=\Delta y^2 + (\frac{m \Delta x}{k_o})^2
 + (\frac{y}{2 k_o \Delta x  })^2 $.  
As expected, the distribution is centered around the classical time-of-arrival $t_c=x_o m/k_o$.  The spread in $y$ has a term due 
to the initial width $\Delta y$ in clock position y.  The second and third term in $\gamma(y)$ is due to the kinematic spread in the time-of-arrival $1/dE =
\frac{m}{k dk}$ and is given by $ \frac{dx(y)\,m}{k_o}$ 
where $dx(y)^2=\Delta x^2 + (\frac{y}{2m\Delta x})^2$. The y dependence in the width in $x$ arises because the wave packet is spreading as time increases, so that at later $y$, the wave packet is wider.  As a result, 
the distribution differs slightly from a Gaussian although this effect is 
suppressed for particles with larger mass.  

When the clock is extremely accurate ie. $p_o >> k_o$
we have $A_T \sim k\sqrt{\frac{2}{mp}}$. 
\beqa
\rho(y,y)_{x>0} &\simeq&\frac{2N^2}{m} \int_{-\infty}^\infty dk\, dk' 
\int_0^\infty dp\, dp'\, dx 
\frac{kk'}{\sqrt{p p'}} 
g(k)g^*(k')f(p)f^*(p') e^{i(q-q')x+i(p-p')y-\frac{i(q^2-q'^2)t}{2m}} 
\nonumber\\
&\simeq& \frac{4 \pi N^2}{m} \int dk\,dk'\, dp 
\frac{kk'}{m} \sqrt{\frac{k^2+2mp}{p (p+\frac{k^2-k
'^2}{2m})}}
g(k)g^*(k')f(p)f^*(p+\frac{k^2-k'^2}{2m}) 
e^{i(k'^2-k^2)\frac{y}{2m}}
\eeqa
Since $p_o>>k_o$, we can approximate this integral as
\beq 
\rho(y,y)_{x>0} \simeq  \frac{A}{m} \mid \int dk\,  k \, g(k)
e^{-i\frac{k^2y}{2m} }|^2
\eeq
where $A \equiv 4 \pi \sqrt{\frac{2}{m}}N^2 \int \frac{dp}{\sqrt{p}} |f(p)|^2$.
We can approximate $p$ by $p_o$ to take it outside the integrand, giving
\beq
A\simeq \sqrt{\frac{\pi}{m p_o}}\frac{2\Delta x }{ \pi^2}.
\eeq
The final integration over $k$ yields
\beq
\rho(y,y)_{>0} \simeq 
4\sqrt{\frac{k_o^2}{2 m p_o}} 
\frac{{\tilde\gamma}(t_c)}{{\tilde\gamma}(y)}
\frac{1}{\sqrt{2\pi{\tilde\gamma(y)}}}
e^{ -\frac{ (y- t_c)^2}{ 2{\tilde\gamma}(y) }  } 
\label{eq:clockdistac}
\eeq
where the width ${\tilde\gamma}(y)=\Delta x ^2 +  (\frac{y}{2 k_o \Delta x  })^2$ is independent of $\Delta y$  because the kinematic spread in the time-of-arrival $ 1/dE$ is much larger than the spread in the 
position of the clock.  In this limit  we see
two additional factors.  The amplitude decays like $\sqrt{E_o/p_o}$ so that improved accuracy decreases our chances of detecting
the particle.  Also, there is a minor correction of 
$\frac{{\tilde\gamma(t_c)}}{{\tilde\gamma(y)}}$. More energetic particles with faster arrival times are more likely to trigger the clock.

\end{document}